\documentclass{article}

\usepackage{microtype}
\usepackage{graphicx}
\usepackage{subfigure}
\usepackage{booktabs} 

\usepackage{hyperref}

\usepackage[accepted]{tom2023}

\usepackage{amsmath}
\usepackage{amssymb}
\usepackage{mathtools}
\usepackage{amsthm}

\usepackage[capitalize,noabbrev]{cleveref}

\theoremstyle{plain}

\theoremstyle{definition}

\theoremstyle{remark}

\usepackage[textsize=tiny]{todonotes}

\icmltitlerunning{From Cognition to Computation: A Comparative Review of Human Attention and Transformer Architectures}

\begin{document}

\twocolumn[
\icmltitle{From Cognition to Computation: \\A Comparative Review of Human Attention and Transformer Architectures}



\icmlsetsymbol{equal}{*}

\begin{icmlauthorlist}
\icmlauthor{Minglu Zhao}{1}
\icmlauthor{Dehong Xu}{1}
\icmlauthor{Tao Gao}{2}
\end{icmlauthorlist}

\icmlaffiliation{1}{Department of Statistics, UCLA}
\icmlaffiliation{2}{Department of Communication, UCLA}

\icmlcorrespondingauthor{Minglu Zhao}{minglu.zhao@ucla.edu}

\icmlkeywords{Theory of Mind, ToM}

\vskip 0.3in
]



\printAffiliationsAndNotice{}  

\begin{abstract}
Attention is a cornerstone of human cognition that facilitates the efficient extraction of information in everyday life. Recent developments in artificial intelligence like the Transformer architecture also incorporate the idea of attention in model designs. However, despite the shared fundamental principle of selectively attending to information, human attention and the Transformer model display notable differences, particularly in their capacity constraints, attention pathways, and intentional mechanisms. Our review aims to provide a comparative analysis of these mechanisms from a cognitive-functional perspective, thereby shedding light on several open research questions. The exploration encourages interdisciplinary efforts to derive insights from human attention mechanisms in the pursuit of developing more generalized artificial intelligence.

\end{abstract}

\section{Introduction} 
Over the past few years, the field of artificial intelligence (AI) has experienced a significant transformation with the introduction of the Transformer architectures, which have rapidly become the cornerstone of many state-of-the-art models in natural language processing (NLP), computer vision, and beyond. These architectures incorporate the concept of attention which echos the complex cognitive process of human attention, a remarkable ability that enables us to focus on specific aspects of our environment while effectively filtering out extraneous information. Our review article aims to provide an in-depth comparative analysis of human attention and Transformer architectures, systematically examining the similarities and differences from various perspectives including vision, language, and agency. 

Attention has been one of the most studied topics in cognitive psychology and influences a broad range of cognitive processes, contributing to perception, memory, and cognitive control. Classic studies in cognitive psychology have underlined attention's role as a filtering mechanism that selectively processes relevant information from the environment while managing cognitive resources \cite{broadbent2013perception, treisman1964monitoring, tanenhaus1995integration, wolfe2000visual}. Additionally, attention as a mental construct is intertwined with self-regulation and social communication processes, facilitating not only individuals' task commitment \cite{rueda2005development, mcclelland2007links, kanfer1989motivation} but also cooperative interactions \cite{bratman1987intention, tomasello1995joint}. On the other hand, AI models with the notion of attention, exemplified by Transformer architectures, have shown great versatility and robustness in various applications. Initially designed for language processing, Transformer models have profoundly transformed the NLP landscape. By utilizing self-attention mechanisms to grasp the contextual relationships in sequences, they have made remarkable advancements in various tasks \cite{brown2020language, openai2023gpt}, including machine translation \cite{vaswani2017attention}, sentiment analysis \cite{devlin2018bert}, and text summarization \cite{liu2019text}. In the computer vision domain, Transformer models have been applied to large-scale visual tasks and are able to capture long-range dependencies and hierarchies in image data \cite{dosovitskiy2020image, carion2020end, khan2022transformers}. Several recent architectures further incorporate Transformer architectures in decision-making tasks, treating decisions as a sequence generation task and thereby enabling the model to learn optimal action strategies based on past experiences \cite{chen2021decision, janner2021offline, meng2021offline}.

The shared terminology of ``attention" in human cognitive studies and Transformer architecture has given rise to intriguing parallels yet certain ambiguities concerning their relationship. Although both emphasize the selective processing of contextual information, human attention and the Transformer model present distinct disparities in their capacity constraints, attentional pathways, and intentional mechanisms. In this article, we systematically compare the two domains around these functional attributes. 
We note that this article does not serve as an exhaustive review of either the human attention mechanism or the Transformer architecture, but rather a focused comparative analysis seeking to identify open directions for better incorporating insights from human attention to attention-based models in AI. We would also like to point out several reviews for a more comprehensive review of the topics \cite{wolfe2000visual,carrasco2011visual, khan2022transformers, guo2022attention}.

\section{Attention modeling and Transformer architecture}
The principle of attention has been incorporated into various domains of AI, underpinning advancements in areas such as image recognition \cite{xu2015show}, speech recognition \cite{chorowski2015attention}, and sequence-to-sequence prediction models \cite{bahdanau2014neural, luong2015effective}. However, in recognition of the significant impact it induced, this paper will concentrate on analyzing the Transformer architecture \cite{vaswani2017attention}, a model that heavily relies on the attention mechanism. The Transformer is an innovative AI architecture and has been a game-changer in natural language processing which has laid the foundation for numerous state-of-the-art models. The Transformer model leverages self-attention mechanisms to capture contextual information in input sequences, making it highly effective in handling long-range dependencies while discarding the use of recurrence in the network. 

\paragraph{Self-attention mechanism}
In essence, the Transformer architecture employs an ``attention" mechanism that takes into account different words' relevance in a sentence when processing each word. This approach can be compared to how we, as humans, selectively focus on certain aspects of a scene or conversation while tuning out the rest. Specifically, this is achieved through the self-attention mechanism, which computes the relationships between all pairs of elements in an input sequence. The core idea is to assign different weights to different elements based on their relevance to the current element being processed. The self-attention mechanism is defined as follows:

\begin{equation}
  \mathrm{Attention}(Q, K, V) = \mathrm{softmax}\left(\frac{QK^T}{\sqrt{d_k}}\right)V,
\end{equation}

where $Q$ (query), $K$ (key), and $V$ (value) are derived from the input matrix $X$ through linear transformations using weight matrices $W^Q$, $W^K$, and $W^V$: $Q = XW^Q, K = XW^K, V = XW^V$. 

The keys ($K$), queries ($Q$), and values ($V$) in this formula can be thought of as abstractions that the Transformer uses to represent different aspects of the words in a sentence. The process of generating key ($K$) and query ($Q$) can be intuitively understood as the Transformer ``asking questions" about certain words (the queries) and ``looking up answers" in other words (the keys). This ``questioning" and ``answering" process allows the Transformer to understand the interdependencies between words in a sentence. The relevance between one word's key and another word's query is computed by taking the dot product of the respective query and key ($QK^T$), which is a measure of their similarity. The $\mathrm{softmax}$ function transforms these relevance scores into probability values that sum up to one. $\sqrt{d_k}$ is a scaling factor that helps the model to be more trainable and stable. Values ($V$) are representations of the words themselves. After the Transformer computes the similarity between all queries and keys, it uses this information to weigh the importance of each value and creates a weighted sum of the values, which forms the output of the attention mechanism. In a way, values are the ``content" that the Transformer wants to focus on, and the role of the keys and queries is to determine how much attention each value should be given.

\paragraph{Multi-head attention}
The multi-head attention mechanism in the Transformer model is a crucial step that allows the model to capture different aspects of the meaning of a sentence. Essentially, multi-head attention means running not one, but multiple attention mechanisms (or ``heads") in parallel, each focusing on different ``types" or ``aspects" of the input sequence's information. This is defined in the following equations:

\begin{equation}
\mathrm{MultiHead}(Q, K, V) = \mathrm{Concat}(\mathrm{head}_1, \dots, \mathrm{head}_h)W^O,
\end{equation}

where each attention head is computed as:

\begin{equation}
\mathrm{head}_i = \mathrm{Attention}(QW^Q_i, KW^K_i, VW^V_i).
\end{equation}

Each attention head applies the attention mechanism but with different learned linear transformations (i.e., different weight matrices $W^Q_i, W^K_i, W^V_i$). Because these transformations are different for each head, each head learns to pay attention to different features in the input. After all the attention heads have processed the input sequence independently, the results from each head are concatenated and then linearly transformed using another learned weight matrix $W^O$. This step integrates the different ``perspectives" from each head into a unified representation, which is then passed onto the next layer in the Transformer model.

\section{Comparative analysis of human attention and attention in Transformers}

We structure our review around key functions of attention as understood from a cognitive perspective and distinguish these based on their similarities or differences when comparing human attention to the Transformer architecture. We outline a series of influential studies in cognitive science regarding the idea of human attention, from domains including vision, language, and agency. We further list out some recent work built on top of the Transformer architecture for comparison.

\subsection{Similarities}
\subsubsection{Selective attention}
From a functional perspective, human attention mechanism can be considered a selective process which allows us to focus on a particular object or task while filtering out irrelevant or distracting information. This phenomenon is often described as the ``cocktail party effect," referring to our ability to concentrate on a single conversation amidst a noisy environment \cite{cherry1953some}. Broadbent's Filter Model further suggested that humans tend to focus on specific elements while filtering out others \cite{broadbent2013perception}. The Attenuation Model by Treisman modified this understanding by proposing that unattended information is not completely filtered out, but rather ``turned down" or attenuated \cite{treisman1964monitoring}. Later, \cite{deutsch1963attention} suggested that the selection of relevant stimuli happens later in the processing chain, indicating selective attention operates not only on the perceptual level. Several lines of work deepened this understanding through researching on spatial attention, where humans filter out spatially nearby distractors for efficient information processing \cite{eriksen1974effects, posner1980orienting}

Similarly, the Transformer model applies a form of selective attention to process sequences of data \cite{vaswani2017attention}. Through its self-attention mechanism, the Transformer model calculates attention scores, or weights, for each element in the input relative to all others. The computation of these weights is managed by a $softmax$ function, designed such that higher weights are allocated to larger input values, while ensuring that the aggregate of all weights sums up to one. In this way, if certain parts receive a higher attention score, the remaining elements are accordingly assigned a smaller fraction of the attention. The mechanism thus allow the Transformer to focus more on some and less on others, echoing how humans tend to concentrate on some inputs while neglecting others.

\subsubsection{Contextual understanding}

Humans typically interpret sensory inputs within their broader context, both in visual scenes and linguistic utterances, a feature commonly referred to as contextual processing. A series of studies demonstrated the effect of context on object recognition \cite{palmer1975effects}. When an object is presented in a congruent scene (e.g., a loaf of bread in a kitchen), people recognize it more quickly and accurately than when it is in an incongruent scene (e.g., a loaf of bread on a beach). A similar effect was found in linguistics studies, where researchers showed that when listeners hear an ambiguous word, they do not perceive both meanings and then select the appropriate one; instead, they immediately interpret the word in light of the surrounding sentence context \cite{tanenhaus1995integration}. This suggests that context is integrated in real time during language comprehension. In studies on human decision-making, researchers revealed how the phrasing of a problem can drastically change people's decisions, even when the underlying objective information is the same. For instance, people may opt for a surgery when told it has a 90\% survival rate, but reject it when told it has a 10\% mortality rate, illustrating how our decisions are shaped by the context in which information is presented \cite{tversky1981framing}. In this way, humans' interpretation of sensory stimuli is informed not only by the immediate properties of the stimuli themselves, but also by their surroundings.

With a similar idea, the Transformer model is designed to accommodate context in processing sequences of data. This is often regarded as one of the major reasons why Transformer models are competitive in multiple domains. Specifically, the self-attention mechanism considers the entire input sequence when processing each token. This design enables the model to accurately capture long-range dependencies in the data, enriching the representation of each token by incorporating the context provided by all other tokens in the sequence \cite{vaswani2017attention}. This design is particularly powerful in tasks such as machine translation and text summarization, where understanding the contextual relationships within the input sequence is critical \cite{vaswani2017attention}. Recent work further take advantage of this design and apply Transformers to sequential decision making tasks, where the action-to-taken is dependent on the context of past experience \cite{chen2021decision, meng2021offline}.

\subsection{Differences}
\subsubsection{Capacity constraints}

One salient distinction between human attention and Transformer models comes from their capacity constraints. The human attention system operates within constraints defined by perceptual and cognitive boundaries, including a limited visual field and working memory capacity \cite{broadbent2013perception, cowan2001magical}. These biological constraints necessitate the deployment of attention to prioritize and extract relevant information. The spotlight of human attention, therefore, shifts and scales based on task demands and environmental cues, enabling efficient processing within these capacity constraints \cite{carrasco2011visual}. The capacity constraint of human attention has been extensively investigated in psychological studies particularly in visual attention. Starting with Broadbent's Filter Model \cite{broadbent2013perception}, it was proposed that human attention serves as a bottleneck that allows only certain information to pass for further processing due to limited capacity. Researchers further introduced the concept of ``selective looking", showing that when engaged in a demanding task, observers can fail to notice unexpected events \cite{neisser1975selective}. In the famous Invisible Gorilla Experiment \cite{simons1999gorillas}, participants watched a video where they were asked to count basketball passes. During the video, a person in a gorilla suit walked through the scene. Despite the conspicuousness of the event, half of the participants failed to notice the gorilla, illustrating the limits of human attention capacity and the phenomenon of inattentional blindness. The Load Theory of attention also suggested that the level of perceptual load in a task determines the efficiency of selective attention \cite{lavie2005distracted}. The capacity constraint is also evident in the visual search tasks, where humans' attentional capacity is guided by feature-based and spatial-based information \cite{wolfe2007guided}. From this perspective, human attention is indeed a double-edged sword: while it facilitates the extraction of needed information from a plethora of sensory input, while simultaneously imposing limits on what can be attended to at any one time \cite{broadbent2013perception, lavie2005distracted}. Moreover, due to the limited capacity of our cognitive resources, human attention typically operates in a sequential manner, focusing on a restricted set of inputs at any given moment and necessitating constant shifts to process diverse environmental elements \cite{broadbent2013perception, dux2006isolation}.

In contrast, Transformer architectures, being artificial constructs, do not possess these inherent biological constraints. Given adequate computational resources, they can process all parts of the input in parallel, regardless of the input's size or complexity \cite{vaswani2017attention}. This parallel processing capability enables Transformers to consider all elements simultaneously, capturing dependencies and relationships across the entire sequence without the need to sequentially shift focus as in human attention. Thus, attention in Transformers can be viewed as a mechanism that enables the model to extract contextual relationships within the data, rather than a solution to a processing limitation.

\subsubsection{Attention Pathways}

Human attention is characterized by a dynamic interplay between top-down and bottom-up processes \cite{corbetta2002control}. Top-down attention is goal-directed and driven by cognitive factors like knowledge and expectations. It allows us to selectively focus on information that is relevant to our current task or intention. This idea is evident in neuroscience studies, where researchers found that top-down influences bias neural representation towards task-relevant information amidst competing stimuli \cite{desimone1995neural}. There is also enhanced sensory activity when anticipating a target stimulus, thus underscoring the priming effect of top-down attention \cite{egner2005cognitive}. From an agency perspective, the top-down attention enables humans to employ various heuristics, achieving more efficient decision-making by directing attention towards relevant or expected stimuli \cite{tversky1974judgment}. On the other hand, bottom-up attention is stimulus-driven and automatic, which is primarily guided by salient stimuli in the environment, such as bright colors, loud noises, or sudden movements \cite{itti2001computational, chun2011taxonomy}. This type of attention helps us quickly identify potential threats or opportunities in our surroundings without the need for conscious control \cite{yantis1998control}. Together, human attention is indeed a dual-pathway mechanism with top-down control guiding the focus of attention, while bottom-up signals determining instant attention shifts \cite{buschman2007top}.

In contrast to human attention, attention allocation in Transformer models is an entirely data-driven process. The attention weights within these models are learned during the training phase, based solely on the input data \cite{vaswani2017attention}. This design paradigm is limited to one direction of information processing, wherein the model aggregates lower-level features to construct higher-level, semantically rich representations. Consequently, the attention focus within Transformers is shaped purely by previous learning experiences, as opposed to the intricate interplay of cognitive factors observed in humans \cite{corbetta2002control}.

\subsubsection{Intentional nature}

In this section, we compare the intentional nature of attention in humans and the Transformers architecture. For humans, the attention mechanism is not solely a passive response to limited cognitive resources, but also a deliberate, controlled process. The process of allocating attention is an active decision-making mechanism, reflecting an individual's agency in consciously directing their cognitive resources in alignment with their beliefs, goals, and intentions, in line with Theory of Mind \cite{bratman1987intention}. The ability to intentionally ignore extraneous information and focus on achieving one's goal is critical to human decision-making. Effective self-regulation of attention plays a crucial role in sustaining commitment to tasks \cite{rueda2005development, mcclelland2007links}, especially for those requiring prolonged effort, such as complex problem-solving and learning \cite{kanfer1989motivation, muraven2000self}, as well as in tasks where multiple outcomes are desirable \cite{cheng2022intentional}. 

Furthermore, attention is a mental construct that significantly contributes to effective communication by serving as a crucial signal for interpreting human intention, emotion, and personality, often indicated by eye gaze. This nonverbal cue provides insightful information into an individual's thoughts and focus. Evolutionarily, humans have uniquely developed a high color contrast between the white sclera and the colored iris, unlike chimpanzees, facilitating discernment of gaze direction and, subsequently, intention and emotions \cite{kobayashi1997unique, tomasello2007reliance}. This visibility of human eye gaze communicates our attention and intention nonverbally with impressive efficacy. Humans, even at a young age, are highly sensitive to the gaze direction of others, enabling an understanding of the intentions and mental states of the person they are interacting with \cite{hood1998adult, tomasello1995joint}. This sensitivity to gaze direction often culminates in gaze alternation between the object of interest and the communication partner, signalling an intention to share attention, thereby inviting social interaction \cite{tomasello2005understanding}. Critically, such shared attention underpins various social interactions and is instrumental in the development of language, social cognition, and theory of mind. Hence, attention operates as a vital component of social cognition by shaping our understanding of each other' intentions.

On the other hand, although the determination of focus in Transformers depends on the learned attention weights, which resonates with the idea of a controlled process as in human attention allocation, it is crucial to note that the Transformer architecture does not possess inherent cognitive states. The concept of ``attention" within this context is essentially a mathematical construct based on learned data patterns. It remains unclear whether the representations output from the attention mechanism coincides with what one believes to be important from an agency perspective. 

\section{Potential Directions}

Drawing upon the comparative analysis of human attention and attention in the Transformer architecture presented thus far, we now turn our focus to identifying salient open research questions. The objective is to explore the extent to which principles of human attention can guide the modeling of attention in artificial intelligence. Specifically, we would like to note that not all characteristics of the human attention mechanism may be desirable when translated into the artificial intelligence domain. Careful judgement is thus required to integrate only those characteristics of human attention that are beneficial in the context of artificial intelligence while avoiding potential impediments. 

\subsection{Is emulating human capacity constraints beneficial?}

Humans possess a limited capacity for cognitive resources including a limited visual field and working memory capacity \cite{broadbent2013perception, cowan2001magical}. These limitations necessitate the need for attention as a cognitive tool that selectively processes information based on its importance. In contrast, Transformer architectures do not face such limitations and handle large volumes of data simultaneously. Despite the disparity, it is important to note that the absence of capacity constraints in Transformers does not denote a deficiency. In fact, one of the Transformer's crucial strengths lies in its ability to process large amount of information in parallel, which has proven critical in recent large-scale models like GPT \cite{brown2020language, openai2023gpt}. Such models have demonstrated impressive performance across a wide range of tasks by leveraging their capacity to parse and process massive corpora of textual data. This computational efficiency is indispensable for AI to effectively aid humans in various fields, such as autonomous driving, where it is crucial to augment human capabilities and overcome cognitive limitations.

\subsection{How can models adopt a resource-rational approach from human attention?}

From another perspective, while the Transformer architecture is not bounded by the same cognitive resource limitations as humans, they may still glean invaluable insights from understanding the efficiency with which humans allocate and utilize their limited resources. From a resource-rational perspective \cite{griffiths2015rational, lieder2020resource}, human cognition demonstrates an important optimization between performance and resource allocation that allows us to operate effectively in complex environments despite our cognitive constraints. Although recent developments in large language models (LLMs) have reached great success, training of the models typically demand substantial computational resources \cite{brown2020language, openai2023gpt}. Unlike humans, who can generate near-instantaneous responses to diverse stimuli, these models require significant computation to produce comparable outputs. Hence, the challenge remains for AI researchers: to maximize efficiency rather than to merely increase capacity. Such economical use may thus pave the way for the development of more robust, effective, and resource-friendly models.

\subsection{How can attention modeling generate meaningful representations?}

Human attention comes with inherent cognitive limitations. This is evident in phenomena such as inattentional blindness \cite{simons1999gorillas} and change blindness \cite{rensink1997see}, wherein humans overlook certain aspects of a visual scene. Beyond indicating a limitation, these phenomena indeed highlight humans' ability to discern the semantically significant components within a complicated visual scene. The human visual attention system does not construct an exhaustive, detailed representation of the visual world; rather, it selects and processes only the crucial segments \cite{treisman1980feature, treue2003visual}. This capacity for discernment underscores the impressive human capability to extract meaningful information from inputs, thereby crafting a useful representation. Contrarily, models like Transformers treat processing as a predominantly data-driven operation, and thus whether the preserved representation is semantically meaningful is still an opaque question that relies on subjective judgements. To develop more generalizable AI algorithms, meaningful representation of one's attention could serve as a critical component to make AI more interpretable and thus trustworthy \cite{doshi2017towards}.

\subsection{How can attention be formulated as an explicit component of agency?}

Beyond a functional construct of human perception, attention further serves as a key element of human agency that can be intentionally modulated and controlled. At an individual level, it operates as a self-regulation mechanism to shield us from irrelevant and distracting information \cite{kanfer1989motivation, muraven2000self}. In social interactions, attention functions as a conduit for understanding the intentions of others and facilitating communication \cite{bratman1987intention}. Research in cognitive psychology indicates that joint attention serves as the foundation of human cooperation, forming a shared commitment to a task \cite{tomasello2005understanding, tang2020bootstrapping, tang2022exploring}. Indeed, theories suggest that the primary objective of conversation is to manipulate each other's attention, thereby fostering a shared common ground of information among cooperative agents \cite{stacy2020intuitive, stacy2021modeling, fan2021learning}. In this way, attention extends beyond merely representing an individual's focus; it serves as an external and explicit construct intertwined with one's agency and implies a more profound role in mediating human interactions and shaping social dynamics. On the other hand, although recent studies incorporating the Transformer-based architecture have emphasized learning from past experiences to decide on action strategies \cite{chen2021decision, janner2021offline}, attention in these models primarily serves to calculate the relationship between all pairs of elements in the input sequence, rather than functioning as a mental mechanism to generate intentional behavior. The concept of joint attention, central to human cooperation from a cognitive science perspective, is not explicitly incorporated in recent developments of multi-agent algorithms based on Transformers \cite{meng2021offline, wen2022multi}. As a result, it remains an open question how we can more effectively incorporate the human perspective of attention into AI models. 

\section{Conclusion}

In conclusion, our review drew a comparison between the human attention mechanism and the Transformer architecture, revealing a diverse range of similarities and differences. We hope the analysis can serve to highlight areas where artificial intelligence might draw inspiration from the intricacies of human attention while also keeping the unique strengths of Transformer-based models. 

\bibliography{reference}
\bibliographystyle{tom2023}



\end{document}